\documentclass{PoS}

\title{Anti-neutrino oscillations with T2K}

\ShortTitle{Anti-neutrino oscillations with T2K}

\author{\speaker{Melody Ravonel Salzgeber}
             \thanks{on behalf of the T2K collaboration}\\
        Universit\'e de Gen\`eve (CH)\\
        E-mail: \email{melody.ravonel@cern.ch}}

\abstract{T2K is a long-baseline neutrino oscillation experiment, in which a muon neutrino beam is produced at J-PARC and detected 295 km away at the Super-Kamiokande detector. The T2K experiment observed electron-neutrino appearance in 2012. This observation enables T2K to explore CP violation in the lepton sector by comparing electron-neutrino appearance and electron-antineutrino appearance. Indeed, the number of observed electron neutrino events up to 2012 is, though within statistical fluctuation, larger than the expectation, which suggests maximal CP violation. Since 2013, T2K has been accumulating data with a muon antineutrino beam. If the suggested maximal CP violation is true, electron-antineutrino appearance would be suppressed. The signal is further suppressed by the smaller cross section for antineutrinos compared to neutrinos. Hence the observation of electron-antineutrino appearance is an important next step. Furthermore, the CPT theorem imposes that the muon disappearance rate must be the same for muon neutrinos and muon antineutrinos; therefore the comparison between neutrinos and antineutrinos is a good test of the CPT theorem, or else a probe for new non-standard interactions of neutrinos with matter. We will report the result of the first search for electron-antineutrino appearance in T2K, as well as a new measurement of muon-antineutrino disappearance to compare with muon-neutrino disappearance measurements. }

\FullConference{The European Physical Society Conference on High Energy Physics\\
                 22-29 July 2015\\
                 Vienna, Austria}

\begin{document}

\section{The T2K experiment}
\begin{figure}[tbh]
  \begin{center}
    \includegraphics[width=1\linewidth]{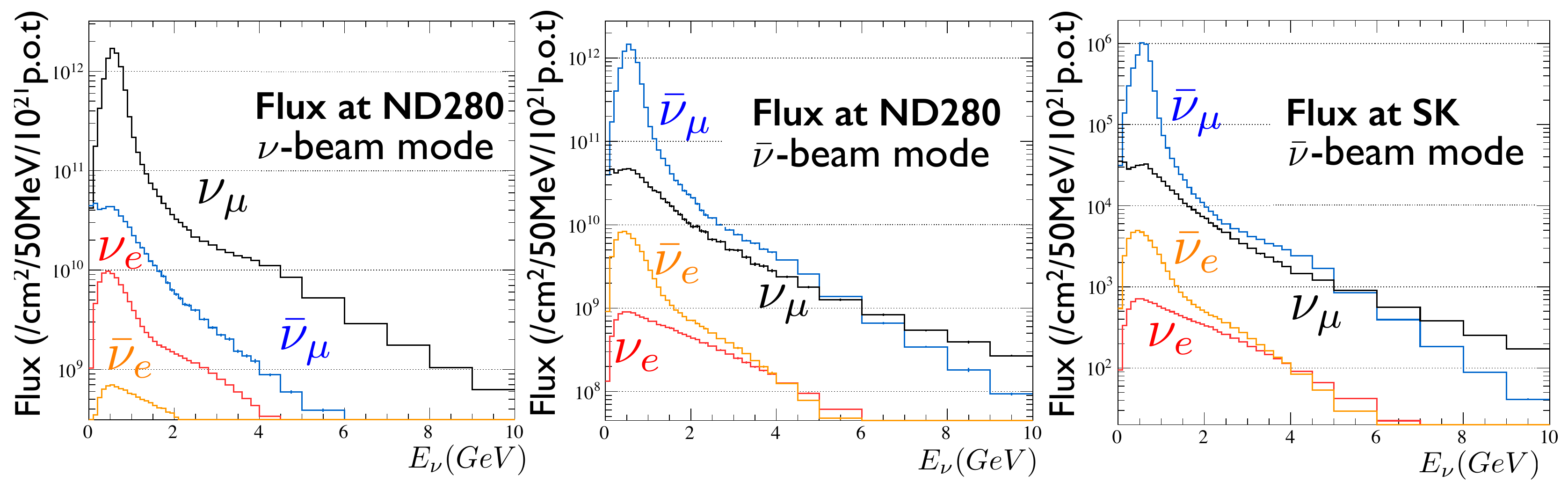}
  \end{center}
  \caption{Left: ND280 off-axis near detector flux prediction with horns operating in +250 kA mode normalized to $10^{21}$ POT. ND280 off-axis near detector (center) and far detector (right) flux prediction with horns operating in -250 kA mode normalized to $10^{21}$ POT. }
  \label{fig:flux_run5c}
\end{figure}

The T2K experiment is a long-baseline neutrino oscillation experiment described in detail in \cite{Abe:2011ks}.
It consists of a far detector at 295~km from the neutrino source and two near detectors at 280~m: an on-axis detector, INGRID, and a 2.5$^\circ$ off-axis detector, ND280, which is used to measure the neutrino or anti-neutrino charged current (CC) rate.
Neutrinos or anti-neutrinos are generated from the 30 GeV J-PARC (Japan Proton Accelerator Research Complex) proton beam located at Tokai-mura, Japan. 
The protons are extracted and interact with a graphite target. The positively chargged (for neutrinos) or negatively charged (for anti-neutrinos) hadrons are focused by three magnetic horns and enter a 96 m long decay pipe, where they decay primarily into muons and muon
(anti)neutrinos. In the case of focusing negatively charged hadrons, mainly muon anti-neutrinos are produced with a relative big contamination of muon neutrinos, as shown in Fig.~\ref{fig:flux_run5c}.

The ND280 near off-axis detector is a multi-purpose set of sub-detectors installed inside a 0.2~T dipole magnet (recycled from the UA1 and NOMAD experiments at CERN) \cite{Abe:2011ks}. 
The tracker is the main component used to constrain flux and cross section parameters in order to decrease the systematic uncertainties in the oscillation analysis. The tracker is composed of three TPCs \cite{Abe:2011ks} and two Fine Grained Detectors (FGDs) \cite{Abe:2011ks}.
For this measurement, the first FGD, which only contains scintillators bars, has been used as active target for neutrino interactions.


The far detector, Super-Kamikande (SK) \cite{Abe:2011ks}, is a 50 kton water Cherenkov detector with very good capabilities to distinguish muons from electrons. 

\section{Near detector samples and fit}
To constrain the flux and cross section parameters, seven near detector data samples are used: three neutrino samples from neutrino beam mode, two anti-neutrino samples from anti-neutrino beam mode, and two neutrino samples from anti-neutrino beam mode. Because the far detector cannot distinguish the charge of particles from a neutrino interaction, it is necessary to understand precisely the various interaction rates at the near detector especially the neutrino contamination in the anti-neutrino beam mode that is one of the main background in the muon anti-neutrino disappearance signal and electron anti-neutrino appearance signal.
The neutrino (anti-neutrino) selections at the near detector rely on the tagging of a FGD1-TPC track compatible with a negative (positive) muon.
While the neutrino samples in the neutrino beam mode are split into three different samples depending on the number of tagged $\pi^+$ (0$\pi^+$,$1\pi^+$,$>1\pi^+$), the other samples are split depending on the number of tracks crossing the TPCs (1 track, $>1$ tracks). 
The total amount of protons on target (POT) used by the near detector fit are $5.8\times 10^{20}$ POT for the neutrino beam mode, and only $0.43 \times 10^{20}$ POT for the anti-neutrino beam mode.

\begin{figure}[tbh]
   \centering
    \includegraphics[width=1\linewidth]{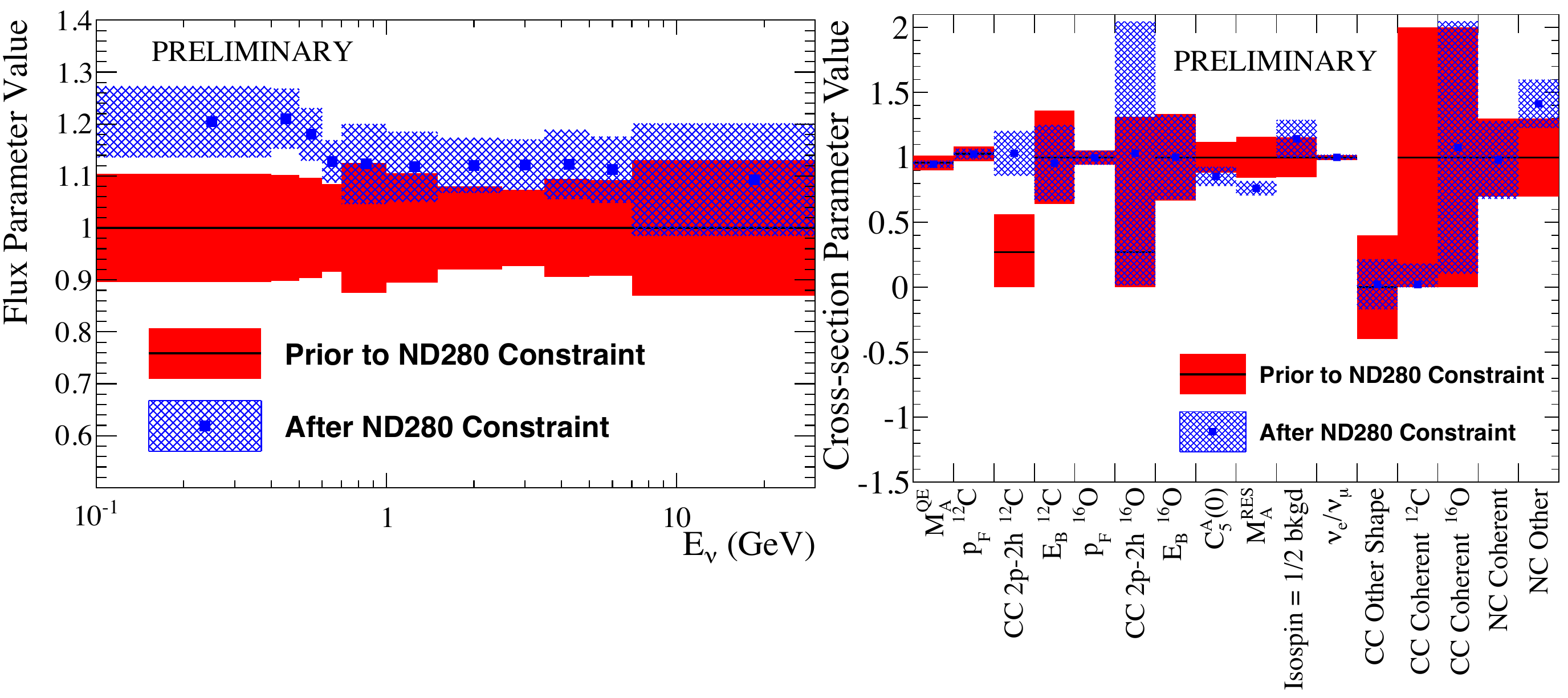}
  \caption{Left: The pre-fit and post-fit SK muon anti-neutrino flux parameters and their uncertainties for the anti-neutrino beam mode. Right: The pre-fit and post-fit cross-section parameters and their uncertainties. Axis labels show the name of each parameter used in the NEUT MC neutrino interaction generator \cite{Hayato:2009zz}.}
  \label{fig:parameters}
\end{figure}

After fitting our Monte-Carlo (MC) to the near detector data, new flux and cross-section parameters are obtained. The uncertainties on these parameters generally decreased from the prior values as we can see in Fig.~\ref{fig:parameters}.  A detailed description of most of these parameters can be found in \cite{Abe:2014ugx,Abe:2015awa}.

The neutrino event generator, NEUT, used for this analysis contains improvements from the generator used in previous results \cite{Abe:2014ugx,Abe:2015awa}. In particular, it contains a first implementation of the multi-nucleon effect on carbon and on oxygen (2p2h in the figure) \cite{Nieves:2013fr}.
It is the first time that this effect is included in a generator for neutrino oscillation analyses. To be conservative, no correlation between Carbon and Oxygen has been taken into account in this first analysis. As a consequence, the reduction of the total uncertainty on the muon anti-neutrino disappearance results is not big and passes from 14.4\% without the ND280 fit to 11.6\% with the use of the ND280 fit results. Once correlation will be taken into account, the systematic uncertainty reduction is expected to be larger at the far detector. Furthermore, we expect an additional reduction as soon as the second FGD containing water is used.


\section{Far detector results}

\begin{figure}[tbh]
   \centering
      \includegraphics[width=0.32\linewidth]{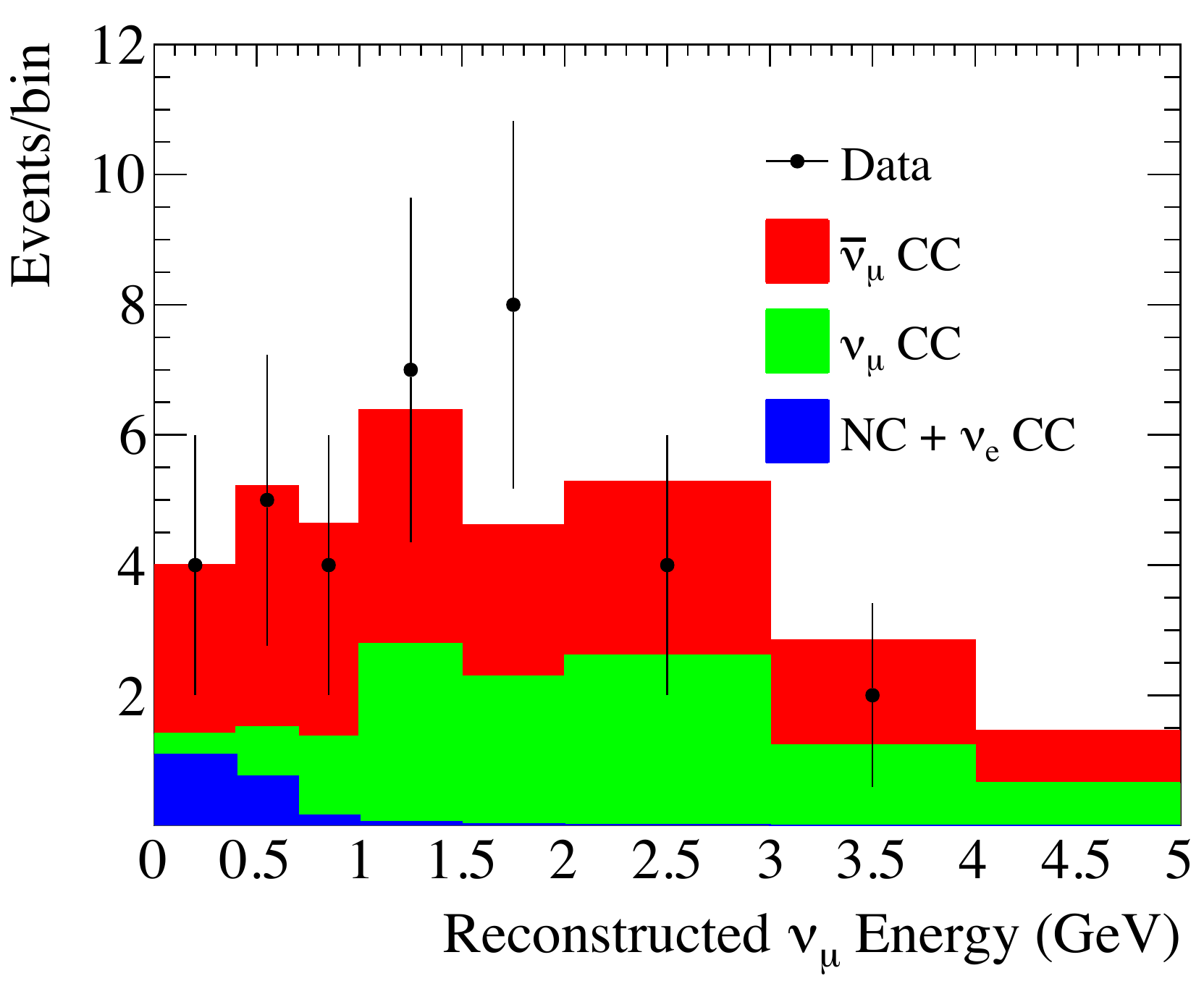}
      \includegraphics[width=0.32\linewidth]{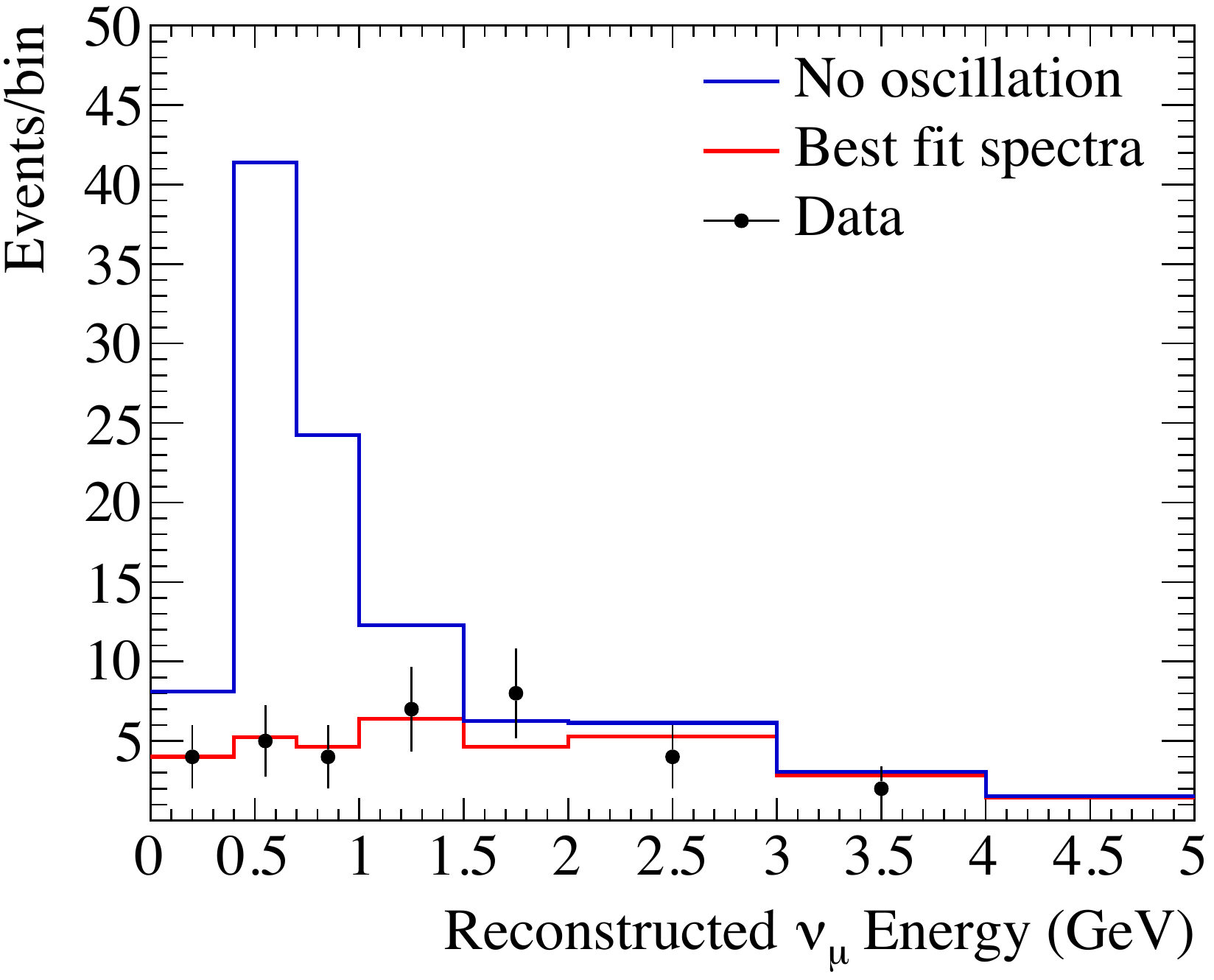}
         \includegraphics[width=0.32\linewidth]{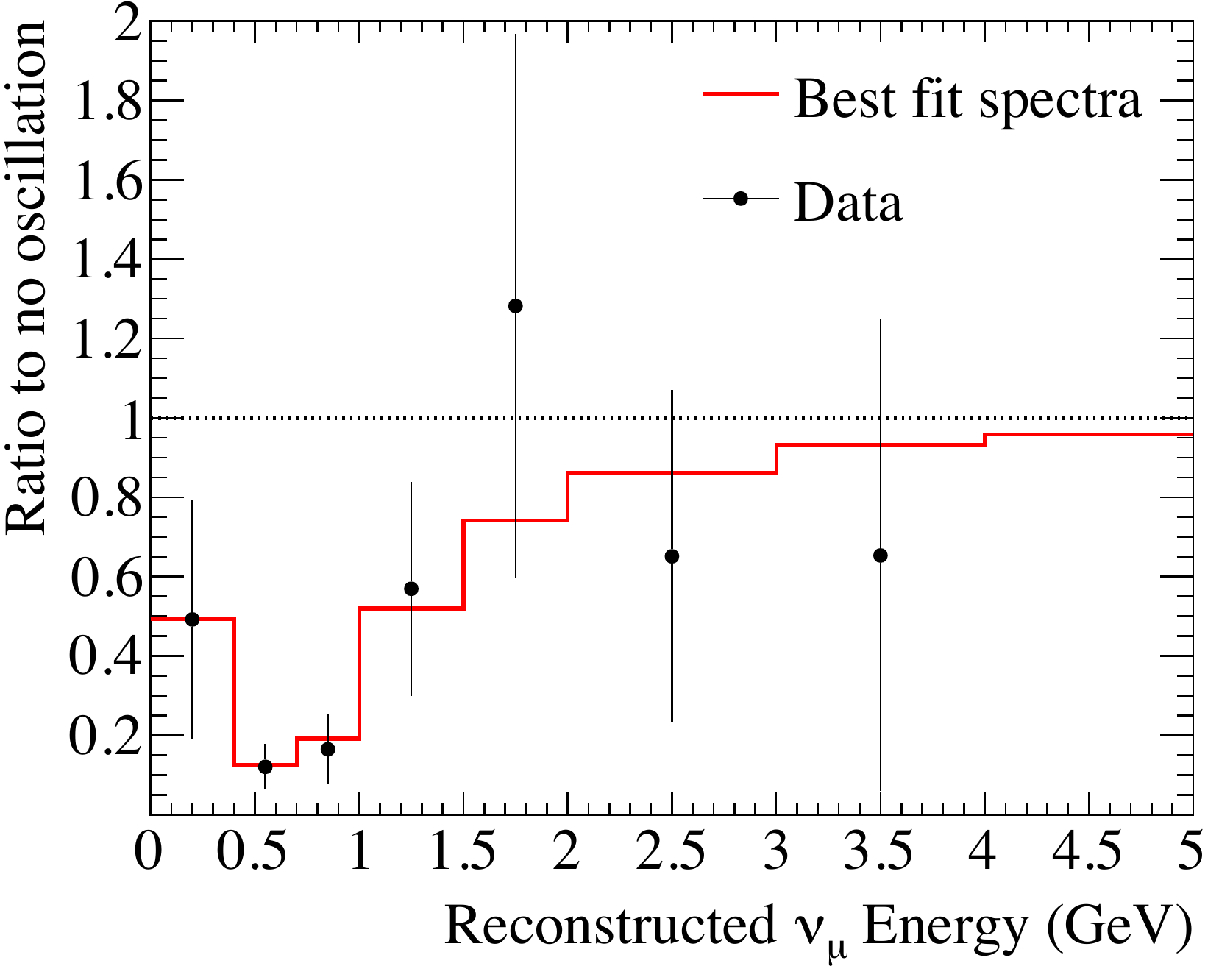}

  \caption{Left: Best fit spectra as a function of the reconstructed energy bin (black dots) together with the expected muon anti-neutrino signal in red, the muon neutrino contamination in green, and the remaining background including neutral current and electron neutrino and anti-neutrino events in blue. Center: Best fit spectra as a function of the reconstructed energy bin together with the un-oscillated spectra. Right: Ratio of the best fit spectrum to the predicted energy spectrum in the case of no oscillations.}
  \label{fig:osc}
\end{figure}

\begin{figure}[tbh]
 \centering
      \includegraphics[width=0.32\linewidth]{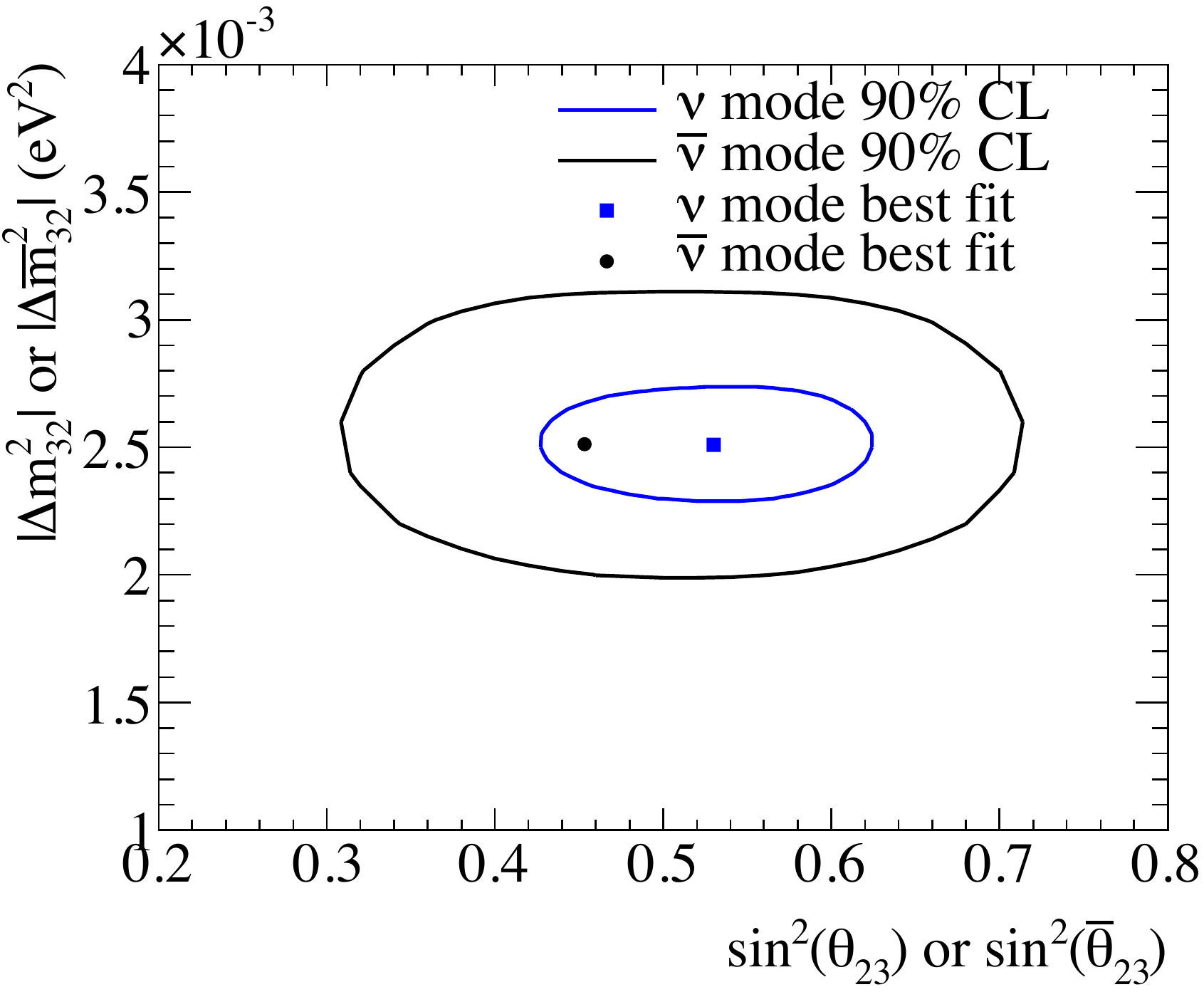}
 \includegraphics[width=0.32\linewidth]{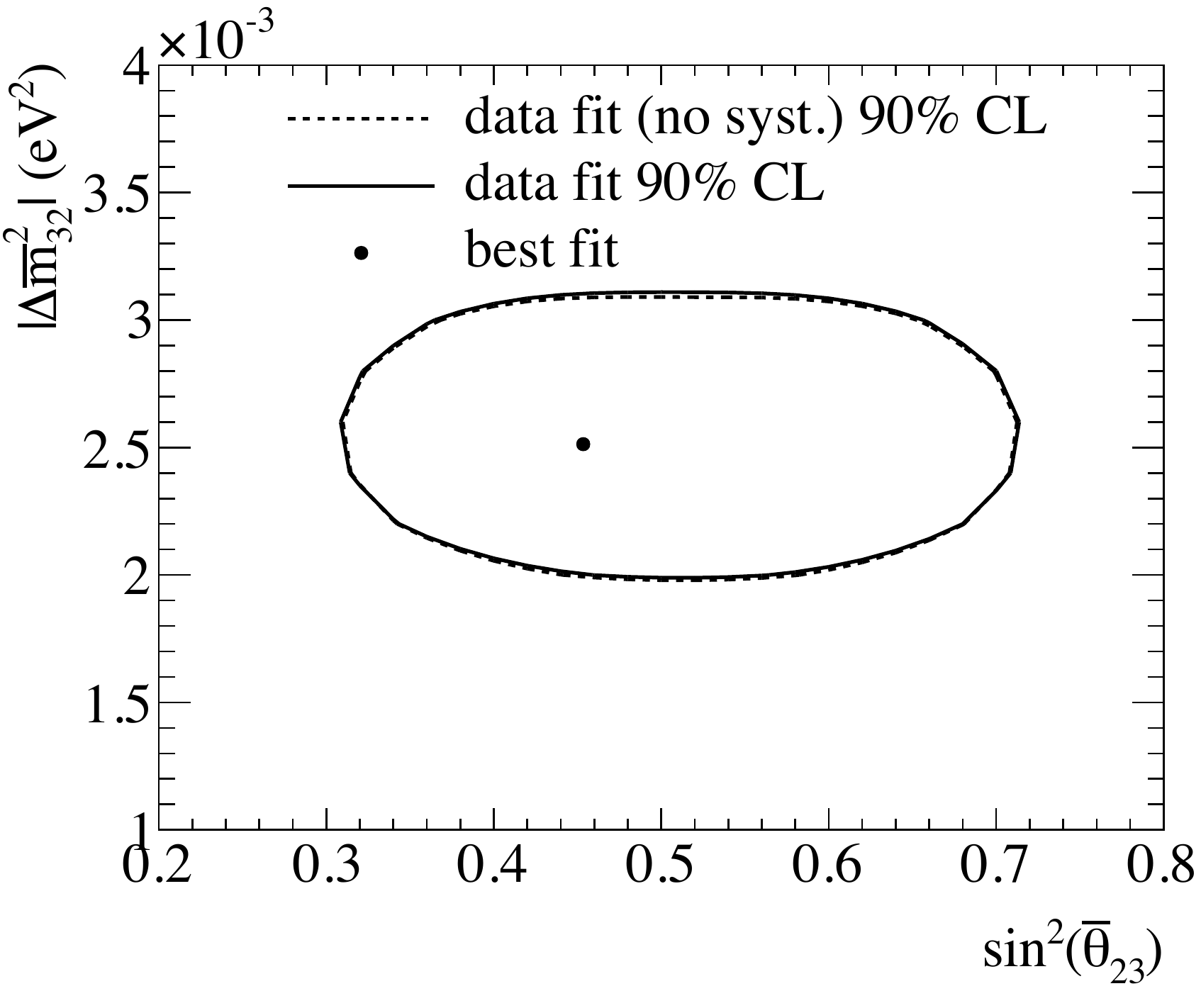}
   \includegraphics[width=0.32\linewidth]{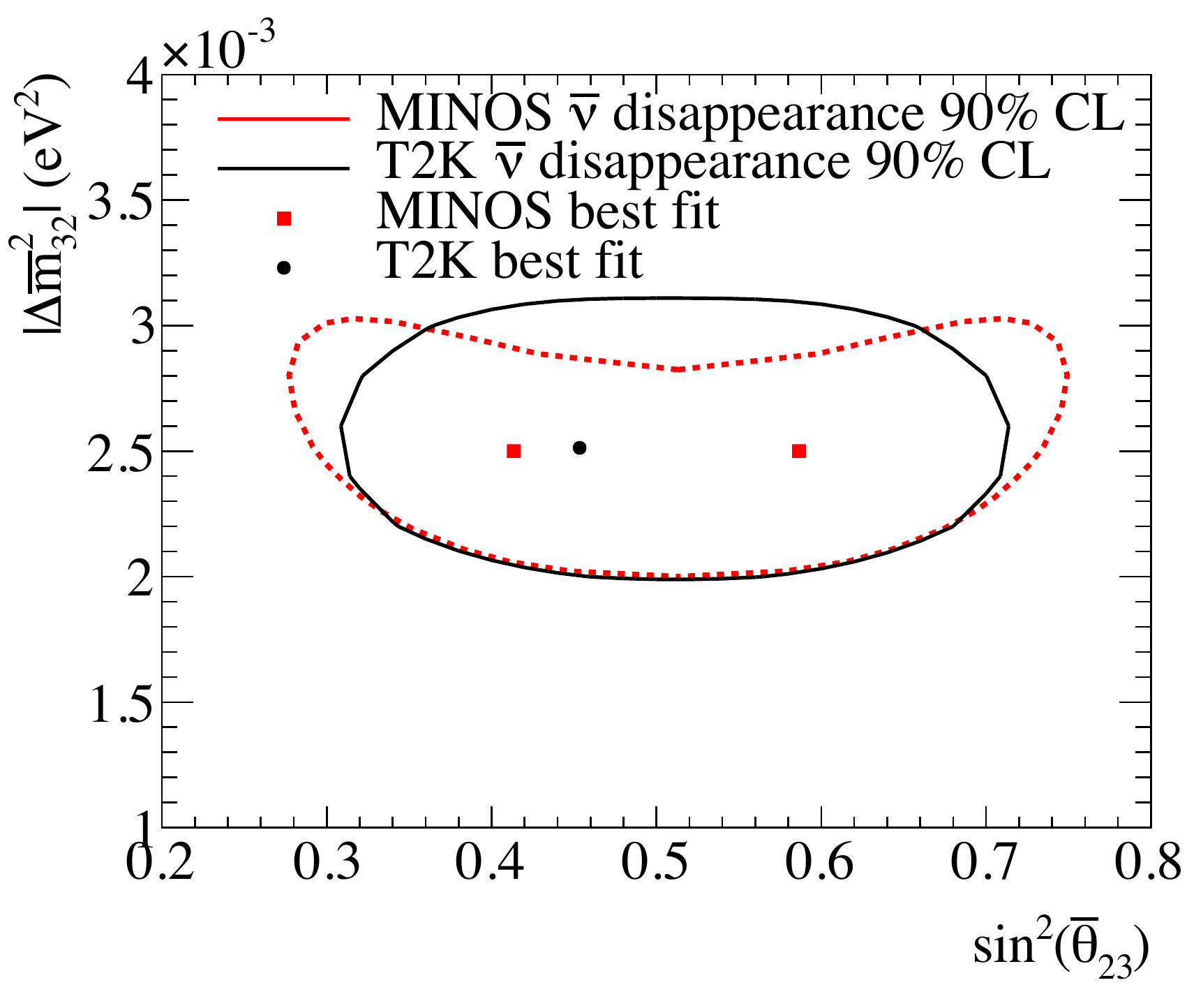}

   \caption{Left: Best fit results compared to the T2K neutrino results \cite{Abe:2014ugx}. Center: Best fit results with and without systematic uncertainties taken into account. Right: Best fit results compared to MINOS results \cite{Adamson:2014vgd}. }
  \label{fig:contours}
\end{figure}

 The far detector selection has been able to use all the POT delivered up to June 2015, which corresponds to $4.04 \times 10^{20}$ POT. This corresponds to 34 single ring muon-like events and 3 single ring electron-like events.
From this data, we see a clear signal of muon anti-neutrino disappearance in Fig.~\ref{fig:osc}  which is still statistically dominated as we can see in Fig.~\ref{fig:contours}. From this figure, we see that anti-neutrino disappearance results are in agreement with MINOS \cite{Adamson:2014vgd} and previous muon neutrino disappearance results \cite{Abe:2014ugx}.

The electron anti-neutrino dataset is statistically limited, and therefore requires a careful analysis. The number of events with electron anti-neutrino is expected to be between 3.7 and 5.5 for this amount of POT. The maximum number of events is obtained for the inverted mass hierarchy and CP phase, $\delta_{CP}=\pi/2$. On the contrary, the minimum number of events is obtained in case of normal mass hierarchy and $\delta_{CP}=-\pi/2$, which is in agreement with the results obtained from the joint three flavour analysis of T2K \cite{Abe:2015awa}. Both numbers are obtained with the following settings:
$\sin^2 \theta_{12} = 0.306$, 
$\Delta m^2_{21} = 7.5 \times 10^{-5} ~eV^2$,
$\sin^2 \theta_{23} = 0.533 $,
$\sin^2 \theta_{13} = 0.0255$, 
$\Delta m^2_{32} = 2.556 \times 10^{-3} ~eV^2$.

\begin{figure}[tbh]
 \centering
    \includegraphics[width=0.9\linewidth]{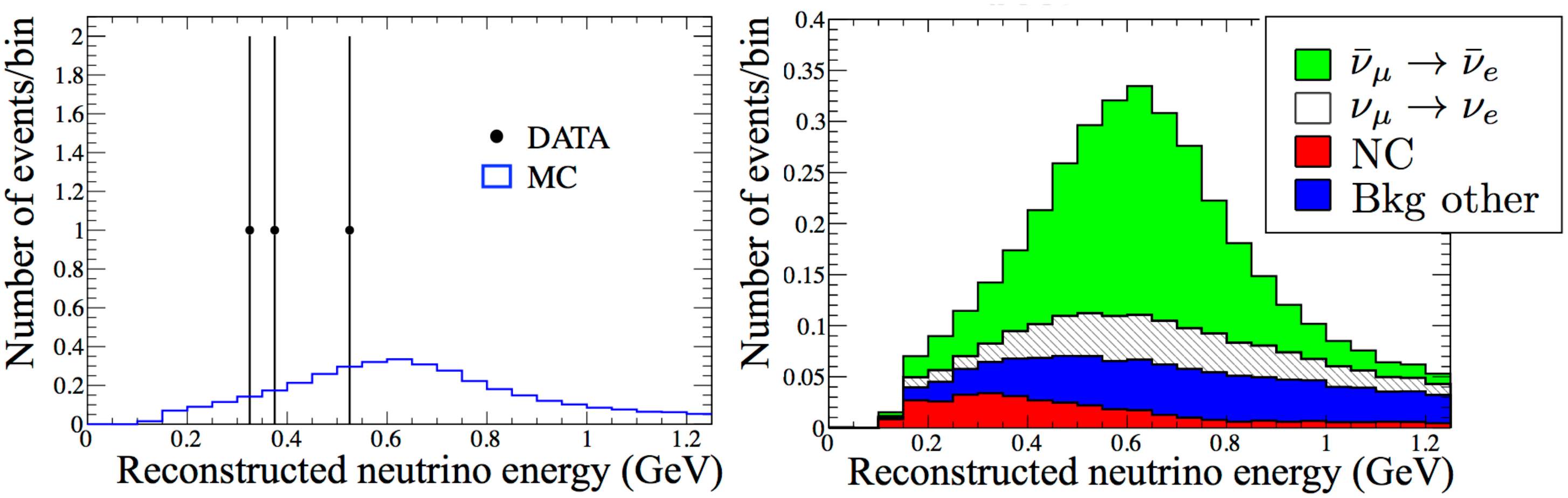}
   \caption{Left: Data point together with the expected MC distribution as a function of the reconstructed neutrino energy. Right: Expected neutrino energy distribution with the MC breakdown.}
  \label{fig:erec_nue}
\end{figure}

\begin{figure}[tbh]
 \centering
    \includegraphics[width=0.9\linewidth]{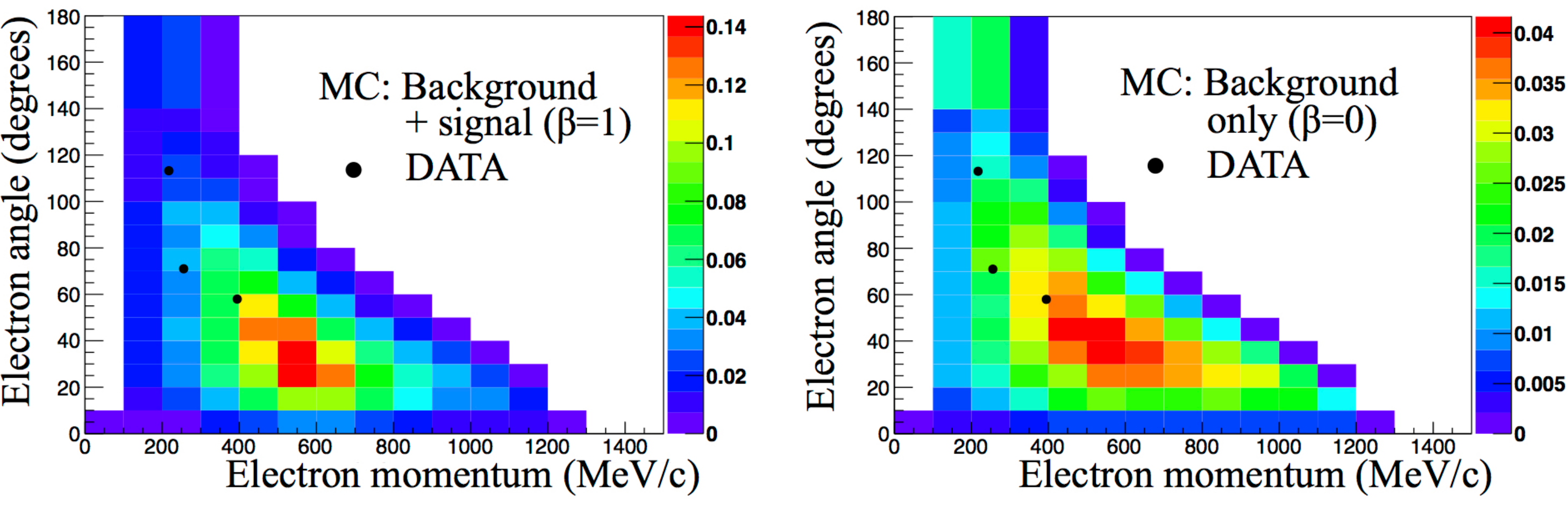}
   \caption{Left: Data point together with the expected distribution as a function of the electron momentum and angle. The filled histograms represent the total number of expected events at Super-Kamiokande including the appearance signal and background. Right: Data point together with the expected background-only distribution as a function of the electron momentum and angle. }
  \label{fig:ptheta}
\end{figure}

Fig.~\ref{fig:erec_nue} shows the distribution of the single ring electron-like events measured at the far detector as a function of the reconstructed neutrino energy and with the breakdown of the background. Fig.~\ref{fig:ptheta} shows two-dimensional distribution of the MC with the data superimposed as a function of the electron momentum and angle. 
From these two figures, we can see that it is difficult to tell if any of the data points can be associated to electron anti-neutrino appearance.
To quantify this observation, we test the no-appearance hypothesis and report p-values and Bayes factors for the current electron anti-neutrino results.

\vspace{-0.1cm}

In this approach, we introduce a discrete parameter $\beta=\{0,1\}$ to modify the electron anti-neutrino probability: if $\beta=0$ then no electron anti-neutrino appearance is taken into account in the calculation of the expected number of events at the far detector, while if $\beta=1$, it is taken into account using the same oscillation parameters as electron neutrino appearance.
The p-value is used to characterize how anomalous our data is with respect to the no appearance hypothesis ($\beta=0$), while the Bayes factor ($B_{10}$) characterizes how our data favors the appearance hypothesis ($\beta=1$) compared to the no appearance hypothesis ($\beta=0$).
We calculate both quantities using, first, only the total number of events at the far detector (rate-only p-value and Bayes factor) and secondly including as well the shape of the reconstructed neutrino energy and the reconstructed out-going electron angle and momentum.

To test our dataset against the null hypothesis, we generate an ensemble of test experiments with $\beta=0$. The rate-only p-value is then defined as the fraction of test experiment in the ensemble with $\beta=0$ that have as many or more candidates as the measured number of events in the Super-Kamiokande detector. The Bayes factor is defined, in this case, as the probability density ($\rho$) ratio between the $\beta=1$ to the $\beta=0$ hypothesis for the measured number of events at the far detector: $B_{10}=\frac{\rho(\beta=1, N_{obs}=3)}{\rho(\beta=0, N_{obs}=3)}$, since we have observed only 3 single ring electron-like events at the far detector.
We obtain a rate-only p-value of 0.26 and a Bayes factor of 1.1 for data. This results are comparable to the expected sensitivity obtained by using an ensemble of fake datasets with $\beta=1$. From the rate only studies, we therefore conclude that our statistics is too small and therefore have little power to reject the background-only hypothesis.

To increase the separation of the test statistic distributions for the two hypotheses, we include the shape information into the analysis, using two different binnings: reconstructed neutrino energy bins, and reconstructed out-going electron and angle bins.
To calculate the \textit{rate+shape} p-value and Bayes factor, we build a likelihood that depends on the number of observed and expected events at the far detector ($N_{obs}$ and $N_{exp}$ respectively) and is defined as:
$$ \mathcal{L}(\beta,\vec{o},\vec{f})=\prod_i^{\mathrm{N~bins}}\mathcal{L}_{\mathrm{poisson}}(N^{obs}_i,N^{exp}_i(\beta,\vec{o},\vec{f}))\times\mathcal{L}_{\mathrm{syst}}(\vec{f})$$
where $i$ denotes the energy bin or the momentum and angle bin, $\beta$, $\vec{o}$ and $\vec{f}$ are the oscillation parameters and underlying systematic parameters, $\mathcal{L}_{\mathrm{poisson}}$ is the \textit{rate+shape} term of the likelihood which follows a Poisson distribution and $\mathcal{L}_{\mathrm{syst}}$ is the systematic term of the likelihood.
The likelihood is then marginalized on oscillation and systematic parameters except $\beta$. The p-value is calculated for the test statistic $-2 \Delta \mathrm{ln}\mathcal{L}=-2 [\mathrm{ln}\mathcal{L}_{marg}(\beta=0)-\mathrm{ln}\mathcal{L}_{marg}(\beta=1)]$.
We obtain for data, $-2 \Delta \mathrm{ln}\mathcal{L}^{DATA}_{p,\theta}=-1.16$ and  $-2 \Delta \mathrm{ln}\mathcal{L}^{DATA}_{E_{\nu}}=0.16$, which gives a p-value of 0.34 and 0.16 respectively and a Bayes factor of 0.55 and 1.1 respectively.
We conclude that our dataset cannot strongly favor either of the two hypotheses.

\vspace{-0.1cm}

\section{Conclusion}
The T2K collaboration has presented updated results on the muon anti-neutrino disappearance measurement that is in agreement with MINOS \cite{Adamson:2014vgd} and previous T2K neutrino analyses  \cite{Abe:2015awa}.
We observed 3 single rings electron-like events. With the present statistic, we cannot significantly favour the appearance hypothesis compared to the background-only hypothesis. 
We therefore need more anti-neutrino data. Depending on the $\delta_{CP}$ phase,
 a stronger evidence will be reached in one or two more years of data taking.

\vspace{-0.2cm}
\bibliography{skeleton}

\end{document}